**REVIEW ARTICLE | OPEN ACCESS**

# Decoding the science behind antioxidant – anti-inflammatory nutraceuticals in stroke

Sophie Béraud-Dufour[1], Ilona Legroux[1], Thierry Coppola[1], Patricia Lebrun[1], and Nicolas Blondeau[1*]

Stroke is a leading cause of disability and death worldwide, with ischemic strokes accounting for nearly 80% of cases. Fewer than 5% of patients receive the sole validated pharmacotherapy, intravenous thrombolysis, highlighting the urgent need for novel therapies. Within this landscape, the exploration of natural molecules emerges as a promising avenue, particularly as a means to address limitations associated with conventional drugs. Nutraceuticals, bioactive compounds derived from food sources, offer a compelling prospect for health and wellness. The term nutraceutical reflects their dual potential in nutrition and pharmacotherapy, emphasizing their relevance to both disease prevention and treatment. Interestingly, many were initially recognized as "natural preconditioners", substances that prime the body for protection against stress or damage. In fact, numerous nutraceuticals have been shown to activate protective pathways similar to those triggered by preconditioning across various organs. Among nutraceuticals, omega-3 polyunsaturated fatty acids sourced from plants or fish, along with polyphenols, have emerged as particularly promising. Their consumption has been associated with a reduced risk of ischemic stroke, supported by numerous preclinical studies demonstrating their beneficial effects on cellular components within the neurovascular unit. This review explores the shared protective mechanisms of various nutraceuticals against key drivers of ischemic injury, including excitotoxicity, oxidative stress, apoptosis, and inflammation. By delineating these actions, the review highlights the potential of nutraceuticals as brain preconditioners that enhance neuroprotection, thereby mitigating the impact of cerebral ischemia in both preventive and therapeutic contexts.

**Highlights**
Key stress responses in stroke, such as excitotoxicity, inflammation, and oxidative damage, lead to long-lasting brain injury. Natural dietary compounds like omega-3 fatty acids and polyphenols may enhance intrinsic cellular defense, potentially reducing stroke risk and severity. Nutraceuticals align with broader conditioning strategies to enhance tissue resilience to ischemia and thereby show promise as low-risk, biocompatible adjuncts to conventional therapies.

## Introduction

The idea behind the review is to inquire about the scientific truths or realities that underlie the attractiveness or appeal of nutraceuticals with antioxidant properties in the context of stroke. In other words, it seeks to examine the scientific basis or evidence supporting the use of nutraceuticals (food or food products that provide health benefits) that have anti-inflammatory, antioxidant, and/or other protective properties in relation to preventing and/or managing strokes. Through several cherry-picked examples, the review aims to understand the connection between nutraceuticals and their potential impact on stroke from a scientific perspective.

## Stroke

Stroke, a dreaded and incapacitating disease, afflicts a staggering 20 million individuals annually, solidifying its position as the second leading cause of death worldwide, with a distressing 30% fatality rate (Benjamin et al., 2019; Campbell et

[1]Université Côte d'Azur - CNRS - Inserm, IPMC, UMR 7275, U1323, Sophia Antipolis, F-06560, France

Correspondence should be addressed to N. Blondeau (blondeau@ipmc.cnrs.fr).

al., 2019). It stands as the primary cause of acquired disability in adults, ranking second only to Alzheimer's disease in dementia cases and following heart disease as the second leading cause of death. The post-stroke sequels frequently encompass paralysis, speech and sensory impairments, cognitive challenges, and enduring disabilities. The repercussions of stroke resonate significantly in the United States, where, in 2016, it contributed to a quarter of all deaths - a disquieting 8.9% surge since 1990. The mortality rate within a five-year post-stroke timeframe is estimated at an alarming 50%. Notably, these somber statistics are accentuated in aging and overweight/obese demographics. Beyond the life-threatening aspects, a substantial 70% of stroke survivors grapple with impaired workability, while 30% necessitate daily assistance for their care. Henceforth, it becomes evident that the pursuit of strategies to mitigate or remedy the deleterious effects induced by stroke is inherently a health imperative. Concurrently, this undertaking assumes economic significance, given the substantial economic burden associated with stroke, estimated at 80 billion dollars in the United States and 33 billion euros in Europe (Benjamin et al., 2019; Campbell et al., 2019).

The challenge inherent in predicting the onset of stroke constitutes a substantial impediment to its preemptive measures. Consequently, the discernment and regulation of predisposing risk factors emerge as imperative prerequisites for the formulation and execution of preventive strategies. Notably, certain factors, including age, sex, and ethnicity, are classified as "non-modifiable" due to their association with inherent and hereditary processes. Illustratively, the incidence of stroke escalates notably from the age of 45, with men exhibiting a higher susceptibility (66% incidence) compared to women (33%). Furthermore, individuals of African or Hispanic lineage manifest an elevated incidence and mortality rate associated with stroke (Allen and Bayraktutan, 2008; Aradine et al., 2022). Conversely, modifiable factors, predominantly entangled with comorbidity and environmental influences, lend themselves to control through lifestyle adjustments, encompassing physical exercise and dietary modifications, or pharmacological interventions aimed at limiting the propensity for stroke. Illustrative instances of these modifiable factors encompass heart disease, diabetes, alcohol misuse, smoking, elevated cholesterol levels, obesity, and hypertension. On one facet, malnutrition characterized by inadequate quantitative intake of vitamins, nutrients, and omega-3 fatty acids not only predisposes individuals to stroke but also constitutes a risk factor for cardiovascular and cerebral pathologies (Dauchet et al., 2005; Riediger et al., 2009; Spence, 2019). Conversely, Western dietary patterns characterized by an excessive abundance of omega-6 fatty acids and an elevated omega-6/omega-3 ratio contribute to the onset of diverse conditions, encompassing cardiovascular diseases, cancer, as well as inflammatory and autoimmune conditions. This was attributed to the conversion of omega-6 fatty acids into pro-inflammatory mediators, such as prostaglandins and leukotrienes. On the opposite, consumption of adequate quantity of omega-3, associated to an optimal omega-6/omega-3 ratio of 4:1 displays protective effects including for brain-related functions (Yehuda, 2003), explaining in part why the Mediterranean diet (rich in polyunsaturated fatty acids [PUFA]), containing fruits, vegetables, olive oil and nuts reduces the risk of stroke in predisposed people (Trichopoulou et al., 2014).

Stroke, more commonly called cerebral attack, predominantly manifests as a consequence of cerebral artery obstruction, accounting for 80% of cases (classified as ischemic stroke), while 20% of cases arise from cerebral artery disruption (identified as hemorrhagic stroke). In the context of ischemic stroke, the occlusion of an artery, denoted as a thrombotic artery, occurs either directly within the central vasculature due to the formation of an athermanous plaque primarily composed of lipids or is instigated by the migration of a thrombus formed extracranially. The latter is subsequently transported to the central circulation, resulting in cerebral embolism (Gorelick, 1989). In both cases, the result is reduced/interrupted irrigation of a more or less large area of the brain, called an infarct, leading to the formation of a hypoperfusion area surrounding the thrombotic vessel and deprivation of glucose, oxygen, and other nutrients necessary for brain survival. The hypoperfusion zone is composed of two regions defined by the severity of reduced blood flow. In the ischemic core, the blood flow is reduced to less than 10 mL/100g tissue/min (Figure 1), whereas at the periphery of the ischemic core, called the ischemic penumbra, the reduction of the blood flow ranges from 10 to 50 mL/100g tissue/min (Fisher and Bastan, 2012). This decreased blood flow and associated ischemia trigger a cascade of deleterious events in all cell types of the neurovascular unit called the ischemic cascade, which has been extensively described in excellent reviews (Dirnagl et al., 1999; Moustafa and Baron, 2008; Woodruff et al., 2011) and briefly recalled below (Figure 2).

At the ischemic core, a ruthless energy deficit occurs, leading to necrosis, an irreversible, premature, and unscheduled death of cells. Indeed, the absence of ATP in this area induces dysfunction of the energy-dependent pump $Na^+/K^+$ ATPase, a key actor in maintaining the electrochemical membrane potential and subsequent uncontrolled sodium influx. This influx is followed by an osmotic water influx, leading to cell swelling, rupture of the plasma membrane, and release of cell contents into the extracellular medium. In contrast to sudden and irreversible death occurring at the ischemic core, cell survival is at first maintained following the occlusion in the ischemic penumbra. However, the viability of neuronal cells later decreases gradually due to the energy deficit, i.e., absence of glucose and oxygen, generated by the disruption of blood circulation. The subsequent shutdown of the resting membrane potential maintenance systems (ATP-dependent) induces a massive depolarization of neurons. Loss of $Na^+/K^+$ ATPase function first triggers plasma membrane depolarization by massive $Na^+$ entry, leading to deregulated release of neurotransmitters, particularly glutamate, which accumulates in the synaptic clefts. The over-activation of glutamate receptors, especially the alpha-amino-3-hydroxy-5-methylisoxazol-4-propionate (AMPA) and N-methyl-D-aspartate (NMDA) receptors, sets off a cascade of events (so called excitotoxicity) of glutamate that ultimately result in cellular damage and death. The over-activation of glutamate AMPA receptors again promotes cell depolarization by bringing $Na^+$ into the cell, removing the $Mg^{2+}$ blockade of NMDA glutamate receptors that allows the massive entry of $Ca^{2+}$ into the cells. The influx of $Ca^{2+}$ is a critical step in excitotoxicity because it triggers various harmful processes within the cell. As an example, the calcium increase activates enzymes such as nitric oxide synthase (NOS) and cyclo-oxygenase (COX) that are responsible for the production of free radicals and various pro-inflammatory molecules, respectively, contributing to cellular damage. Cumulative damage further activates excitotoxic pathways, leading to a self-perpetuating cycle from excessive $Ca^{2+}$ influx, oxidative stress, and inflammation.

The generation of free radicals induces oxidative stress, marked by an equilibrium disruption between the excessive production of reactive oxygen and nitrogen species (ROS/RNS) by mitochondria and the antioxidant capabilities of the cell. The overwhelming of cellular antioxidant defenses impedes their ability to counteract oxidative damage, particularly peroxidation of lipids, proteins, and DNA. Hence, oxidative stress has direct detrimental effects on various types of brain cells. By damaging neuronal membranes and proteins, disrupting cell signaling, it promotes neuronal death. By impairing astrocyte supportive functions, such as neurotransmitter level regulation and blood-brain barrier (BBB) maintenance, it contributes to neuroinflammation and compromises the overall homeostasis of the brain environment. Additionally, it triggers dysfunction in endothelial cells, compromising the integrity of the BBB, which is a significant factor in the post-ischemic inflammation phase, notably facilitating the infiltration of leukocytes to the site of inflammation, contributing to the inflammatory response.

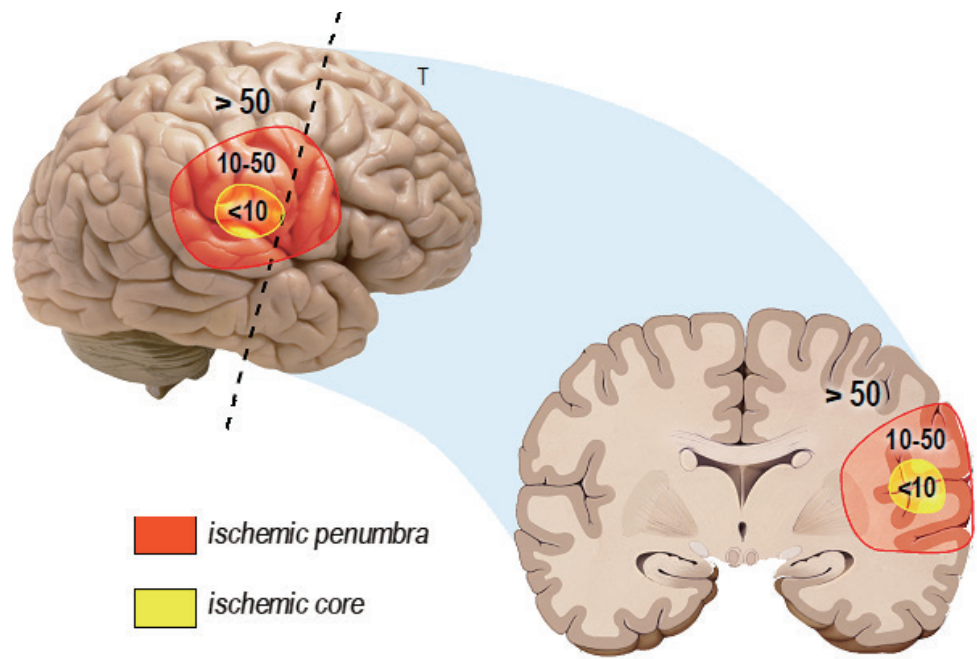

Figure 1. Spatial model of cerebral blood flow reduction in the ischemic area. The figure illustrates the reduction in cerebral blood flow as a function of the ischemic zone, highlighting the hypoperfusion area in mL/100 g tissue/min. The hypoperfusion zone comprises two distinct regions characterized by the severity of blood flow reduction. In the ischemic core, blood flow diminishes to less than 10 mL/100g tissue/min, representing a critical reduction for cell survival. Meanwhile, at the periphery of the ischemic core, known as the ischemic penumbra, the reduction in blood flow varies within the range of 10 to 50 mL/100 g tissue/min, representing a critical reduction in cell functioning. This diminished blood flow and the associated ischemia initiate a cascade of detrimental events across all cell types within the neurovascular unit, collectively referred to as the ischemic cascade.

Finally, oxidative stress induces microglia activation into a pro-inflammatory state through Toll-like receptor 4 receptors (TLR-4 receptors), triggering activation of the inflammatory pathway involving the nuclear factor kappa B (NF-κB) transcription factor in the microglia. NF-κB is an activator of the pro-inflammatory genes of immune cells. Its inactive form is linked to the inhibitor IκB and is sequestered in the cytoplasm. Subsequently, to TLR-4 activation, IκB is degraded, removing NF-κB inhibition. NF-κB is then imported into the nucleus and induces the expression of tumor necrosis factor-alpha (TNFα), interleukin (IL)-1β, and IL-6 that, mainly via TNFα, directly induce death-signaling pathways. Indirectly, pro-inflammatory cytokines induce the activation of peripheral leukocytes and their infiltration into the brain. IL-1β promotes the expression of adhesion factors on endothelial cells lining the BBB, allowing leukocytes to interact with and pass through the barrier to reach the site of inflammation (Huang et al., 2006). The chemokine monocyte chemoattractant protein 1 also guides the migration of monocytes to the site of inflammation and increases the permeability of the BBB, contributing to leukocyte infiltration and thus amplification of the brain lesion (Stamatovic et al., 2005; Le Thuc et al., 2015). Subsequently, activated leukocytes secrete pro-inflammatory cytokines and neurotoxins, such as nitric oxide (NO), and phagocyte damaged neurons. NO production by microglia is also exacerbated by the massive release of glutamate, as well as proteolytic enzyme secretions, such as metalloproteinases (MMP)-9 and MMP-3 that degrade the extracellular matrix of the vascular endothelium, thus breaking the integrity of the BBB and facilitating leukocyte infiltration (Figure 2).

In summary, with the brain functioning on a just-in-time basis for its energy supply, the absence of reperfusion (i.e., the return to normal blood flow) or treatment allows the spread of ischemic damage from the ischemic core to the ischemic penumbra, resulting in massive brain cell death.

Currently, emergency trucks cannot initiate treatment; patients must be transported to a hospital where diagnostic tools, such as X-rays and MRIs, are required. There is no existing drug that prevents stroke, and the primary therapy involves restoring blood flow through thrombolysis (lysis of the blood clot by drug injection) or thrombectomy (mechanical unclogging of the artery). Both interventions must be administered within specific time frames. Thrombolysis is typically administered up to 4.5 hours post-stroke (Etherton et al., 2020). Beyond this period, the risk of adverse effects, particularly bleeding, significantly increases. In contrast, for selected patients, thrombectomy has shown successful outcomes when performed up to 24 hours after a stroke. However, the optimal period for achieving maximal outcomes is within the first hour following the onset of a stroke (Nogueira et al., 2018).

Given these limitations, there is a critical need to explore alternative preventive and therapeutic strategies for stroke. Despite numerous preclinical studies targeting the ischemic cascade, resulting in the identification of potential neuroprotective candidates, none have translated into a viable therapeutic option (O'Collins et al., 2006). Among them, a potent competitive NMDA antagonist, Selfotel (CGS-19755), which is highly protective in animal models, was found to be harmful and responsible for psychiatric and cardiovascular side effects (Perez-Pinzon and Steinberg, 1996; Davis et al., 2000). In order to block calcium entrance into cells, which is a consequence of the excitotoxic phenomenon, additional studies focused on calcium channel blockers, highlighting, for example, the nimodipine molecule. However, this and other calcium channel blockers did not show any benefit in clinical trials (Group, 1992; Zhang et al., 2019). In the area of oxidative stress phase and neutralization of ROS production, additional studies led to the characterization of Tirilazad in particular (a lipid peroxidation inhibitor), but this molecule was found to be without benefit or even harmful during clinical trials (Haley, 1998). Finally, in case of post-ischemic inflammation, Enlinomab, an anti-intercellular adhesion molecule-1 that inhibits leukocyte infiltration, was studied up to the clinical trial phase, but worsened the outcome of treated patients (Enlimomab Acute Stroke Trial, 2001).

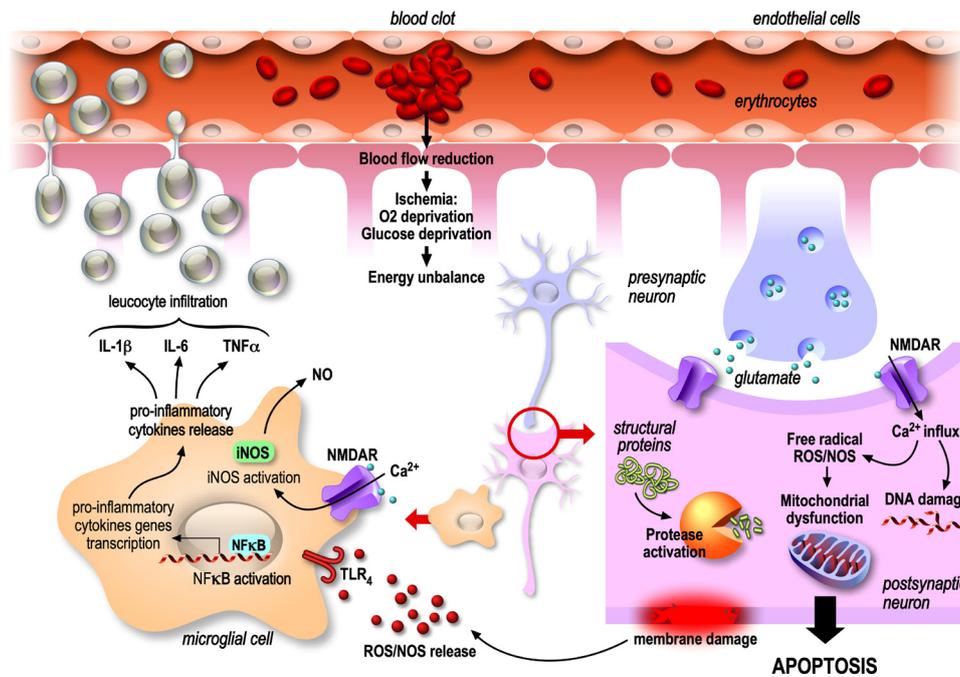

Figure 2. Deleterious mechanisms of ischemic stroke possibly targeted by nutraceuticals. This diagram depicts a longitudinal section of an occluded artery and various deleterious components of ischemic stroke that are potential targets for natural compounds. The artery walls are lined with endothelial cells, and within the vessel, key targets proinflammatory leukocytes surrounded by astrocytes, pre- and post-synaptic neurons, and microglia cells, emphasizing the intricate involvement of different cell types in the pathophysiology of ischemic stroke. The identification of these targets provides valuable insights for potential interventions using nutraceutical to mitigate the impact of ischemic stroke.

A simple and ancient concept would be to opt for the use of natural compounds to avoid the risk of undesirable effects associated with the use of chemical compounds. What's more, the benefits of natural compounds would be even greater if they were already consumed in the daily diet, whether from the plant or animal world, as they would offer greater compatibility with human metabolism, thus reducing the risk of side effects. Recent studies focusing on the intrinsic properties of natural molecules contained in the diet, such as polyunsaturated fatty acids and polyphenols, suggest an increased interest in their potential for stroke treatment (Leng et al., 1999; de Goede et al., 2011; Bayes et al., 2023; Lopez-Morales et al., 2023). Their addition to the diet or their administration could represent a paradigm shift in stroke management (Blondeau and Tauskela, 2013). This study aims to scientifically support the concept that natural molecules derived from nutrition could protect the brain from stroke-related damage and enhance recovery of damaged tissue, and that it could be used to develop new drugs.

### Nutraceuticals - new hope for both prevention and treatment of stroke

Recognized for their cardiovascular and cognitive benefits, certain natural compounds that exhibit antioxidant and anti-inflammatory properties may be worth incorporating into one's daily diet, thereby potentially enhancing overall well-being. Moreover, these properties extend to the preservation of brain health, and their inclusion in the diet is anticipated to be neuroprotective, a term referring to protective strategies and mechanisms tailored for the central nervous system. This involves "all" approaches that preserve neuronal structures and functions in the presence of disorders by inhibiting deleterious cellular mechanisms and reducing irreversible brain damage (Auriel and Bornstein, 2010). Some natural compounds have already been shown to possess neuroprotective properties and are therefore of interest as "nutraceuticals". Originally, this term, which is a combination of nutrient and pharmaceutical, refers to foods or any of their constituents that have beneficial health effects, including the prevention or treatment of disease (DeFelice, 1995). Nowadays, this definition has been redefined and limited to agents isolated from food, whose efficacy could be evaluated in medicinal forms, at least at the preclinical level, and which could be supplemented in the diet. Some nutraceuticals, such as omega-3 alpha-linolenic acid and the polyphenol resveratrol, which have been shown to mimic ischemic preconditioning of the brain (Blondeau et al., 2002; Raval et al., 2006), demonstrated neuroprotective and/or anti-inflammatory potential in in vitro and/or in vivo stroke models (Nguemeni et al., 2013; Silva and Pogacnik, 2020). The development of neuroprotective treatments based on nutraceuticals could provide new therapeutic hope for preventing and slowing down the deleterious mechanisms of stroke (Blondeau, 2016; Tauskela et al., 2017; Chelluboina and Vemuganti, 2021). In this context, the health benefits of PUFAs and polyphenols, which are prevalent in oils, teas, fruits, and vegetables, may serve as exemplary nutraceuticals due to their well-documented health benefits.

### The omega-3 PUFA

#### Alpha linolenic acid (ALA)

PUFAs comprise two or more double bonds in a chain of hydrocarbons. The position of the first double bond defines the PUFA family: omega-3 or omega-6. These fatty acids present an amphiphilic structure, i.e., a hydrophilic head and a hydrophobic tail, and are constituents of the cell membrane, contributing to its fluidity (Wiktorowska-Owczarek et al., 2015). The alpha linolenic acid (C18:3 n-3, ALA) is an

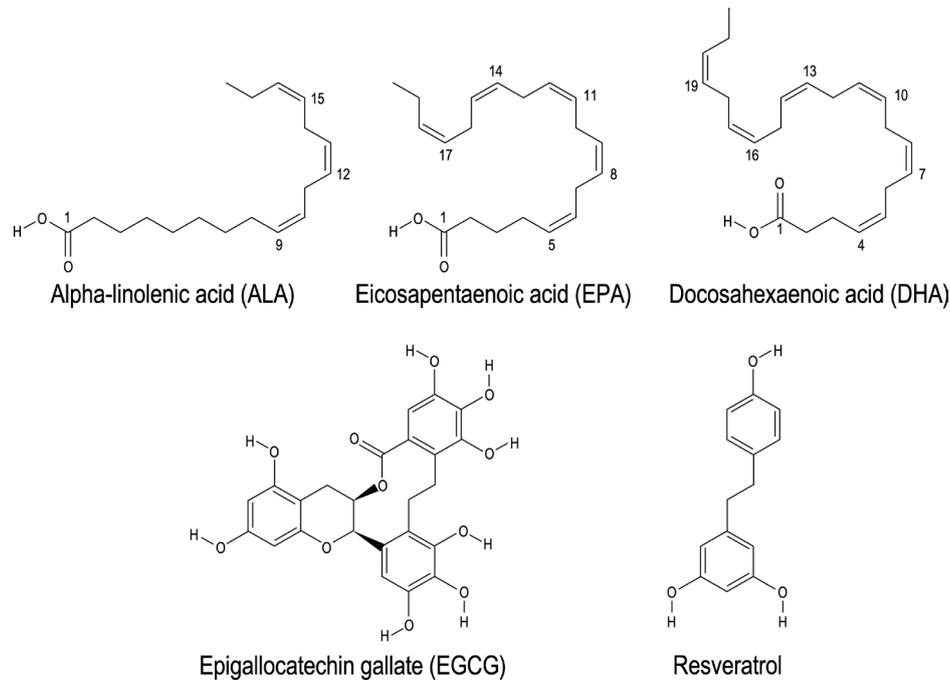

Figure 3. Chemical structure of 2 different kinds of nutraceuticals: The figure illustrates the chemical structures of two distinct groups of nutraceuticals known for their health-promoting properties. The first group comprises omega-3 polyunsaturated fatty acids, including alpha-linolenic acid (ALA), eicosapentaenoic acid (EPA), and docosahexaenoic acid (DHA). The second group features polyphenols, encompassing epigallocatechin gallate (EGCG) and resveratrol. This structural diversity contributes to their biological activities and is linked to their potential positive impacts on health.

omega-3 PUFA, contained mainly in rapeseed, nuts, flaxseed, and rapeseed oils (Figure 3). It is an essential fatty acid, as humans do not synthesize it and it must be obtained through the diet. Experiments performed in both in vivo and in vitro stroke models have demonstrated several of its beneficial effects (Nguemeni et al., 2013).

ALA reduces glutamatergic hyperexcitability induced in the ischemic penumbra.
In glutamatergic neurons, in vitro, the addition of ALA reduces the membrane depolarization induced by the glutamate analogue, kainate. Indeed, by opening TREK-1, a neuronal background potassium channel widely expressed in neuronal brain cells, ALA induces a $K^+$ outward current, hyperpolarizing the plasma membrane, thereby reducing glutamate-induced hyperexcitability (Nguemeni et al., 2013). Moreover, the hyperpolarization also prevents post-synaptic transmission and thus the massive and deleterious entry of calcium. These observations are corroborated by several in vivo studies demonstrating the beneficial effect of an ALA-rich diet on seizures. Mice fed a diet enriched in rapeseed oil showed a decrease in the epileptogenic threshold induced by the injection of kainate, when compared to mice fed a diet enriched in omega-6 (Pages et al., 2011). Moreover, the protective effect of ALA includes various other features. One key aspect is ALA's ability to compete with the omega-6 PUFA arachidonic acid, an essential fatty acid, as a substrate in the biosynthesis of prostaglandins, thromboxanes, and leukotrienes (Calder, 2006), and isoprostanes (Jahn et al., 2008). Additionally, the protective effect may involve the generation of bioactive lipoxygenase-derived omega-3 metabolites, specifically linotrins (Lagarde et al., 2014; Balas and Durand, 2016; Balas et al., 2022), and the non-enzymatic oxygenated metabolites called phytoprostanes (Ahmed et al., 2020).

ALA is beneficial for the vascular compartment.
Application of ALA on the basilar and carotid arteries in mice demonstrates a 30% specific vasodilation of the basilar arteries. The vasodilatory effect induced could be attributed to its impact on the TREK-1 channel found in the myocytes (smooth muscle cells) of the brain arteries. This interaction leads to the relaxation and hyperpolarization of smooth muscle fibers, resulting in arterial dilation. Consequently, ALA exhibits a specific and localized vasodilatory effect that could potentially enhance oxygenation of brain tissue in the event of a stroke (Blondeau et al., 2007). In parallel, potential cardiovascular benefits of ALA have been studied, including its impact on blood pressure regulation. Studies have suggested that oral administration of ALA-rich linseed oil may contribute to lowering systolic blood pressure in hypertensive rats. This effect is thought to be associated with increased concentrations of vasodilating agents, such as bradykinin and nitric oxide (NO) metabolites. ALA's inhibition of angiotensin-converting enzyme (ACE) and its prevention of bradykinin inactivation would therefore exert a systemic vasodilator effect, reducing hypertension, which is a main risk factor for stroke (Sekine et al., 2007; Matsuo and Yagi, 2008).

ALA promotes neurogenesis and synaptogenesis.
Mice subjected to ALA injection exhibit heightened expression of proteins associated with the regulated exocytosis machinery, specifically synaptophysin, SNAP-25, VAMP2, and the glutamatergic transporter V-Glut, within the hippocampus and cortex. Concurrently, elevated levels of the neurotrophic factor brain-derived neurotrophic factor (BDNF) are observed in the hippocampus, striatum, and cortex (Matsuo and Yagi, 2008). BDNF is recognized for its involvement in memory and learning processes, promoting neuronal survival, and potentially fostering increased synaptogenesis. These effects are particularly pertinent to the preservation of neurological functions in regions impacted by ischemia. Further in vivo investigations demonstrate that intravenous administration of ALA in mice, administered prior to the induction of cerebral ischemia through middle cerebral artery occlusion (MCAO), induces a dose-dependent expression of the nuclear factor NF-κB in the hippocampus (Blondeau et al., 2001; Pan et al., 2012). While NF-κB is implicated in inflammatory mechanisms leading to apoptosis, it also possesses the capability to enhance

BDNF expression (Lipsky et al., 2001) offering a multifaceted interplay for ALA in NMDA receptor-mediated neuroprotection in the context of cerebral ischemia (Lipsky et al., 2001).

ALA reduces inflammation and neuroinflammation.
ALA pre-treatment of microglial cells has been shown to attenuate lipopolysaccharide (LPS)-induced expression of inducible NOS (iNOS). Thus, ALA mitigates the toxic production of NO and the subsequent stimulation of brain inflammation. Notably, ALA, administered either as a pre-treatment or post-treatment, exhibits a suppressive effect on the production of pro-apoptotic cytokines, including tumor necrosis factor α (TNFα) and COX-2 (Carey et al., 2020). The latter enzyme, COX-2, plays a pivotal role in prostanoid synthesis, serving as a mediator in the body's inflammatory response. A similar anti-inflammatory impact has been observed in vivo in the context of head trauma (Desai et al., 2016). These outcomes imply that ALA has the potential to regulate inflammation by attenuating the levels of pro-inflammatory cytokines and prostaglandins, thereby mitigating adverse mechanisms associated with stroke.

ALA in Clinical Trials.
Associations between ALA and protection against stroke have been studied in human cohorts. A study conducted in the USA involving 191 middle-aged men with high cholesterol, diastolic hypertension, and a history of smoking established a correlation between elevated circulating levels of ALA and a notable 37% reduction in stroke risk (Simon et al., 1995). However, contrasting findings emerge from a recent Danish study focused on a population aged 50-64 years, where the link between ALA consumption and ischemic stroke appears weak and statistically insignificant (Bork et al., 2018). It is important to note that this study lacked data on ALA concentration in the serum, and ALA intake was assessed using a food frequency questionnaire, making it challenging to assess the physiological impact of ALA consumption. Finally de Goeth's study (de Goede et al., 2011) indicates that, in a general Dutch population, a higher ALA intake leads to a 35-50% lower risk of incident stroke, and lower intake of ALA is a risk factor for incident stroke.

ALA demonstrates pleiotropic properties, positively influencing the entire neurovascular unit. The highlighted results emphasize the protective impact of ALA consumption on neuronal survival and brain inflammation, aligning with its clinical relevance in acute stroke treatment. Notably, ALA's benefits extend beyond direct action, encompassing its metabolites, including eicosapentaenoic acid (C20:5 n-3, EPA) and docosahexaenoic acid (C22:6 n-3, DHA). While physiologically low liver conversion of ALA to EPA (6%) and DHA (3.8%) has been demonstrated, these metabolites are naturally abundant in fish and krill oils, supporting their application as nutraceuticals (Gerster, 1998; Ulven et al., 2011).

**Long-chain PUFAs**

*EPA and DHA*

EPA and DHA have neuroprotective effects in vivo.
For several decades, anti-inflammatory and antioxidant effects have been attributed to EPA and DHA. The significance of such omega-3 fatty acids derived from fish has been described in supporting the body's optimal functioning throughout life (for review see: Swanson et al., 2012). An early study performed by our group showed that both EPA and DHA also have some promising therapeutic potential against cerebral pathologies, mirroring previous propositions for their efficacy in cardiac diseases. They demonstrate the ability to prevent neuronal death in an animal model of transient global ischemia, even when administered post-insult. Furthermore, they provide protective effects against seizures and hippocampal lesions in animals treated with kainite (Lauritzen et al., 2000). Moreover, these PUFAs confer robust tolerance against neurodegeneration. This protective effect was demonstrated in two distinct models involving kainic acid injection and global ischemia, known to induce neuronal death (Blondeau et al., 2002). The neuroprotective and anti-ischemic capabilities of DHA injection were confirmed in a rat model and a mouse model of MCAO (Belayev et al., 2009), as well as with an acute intravenous injection of an emulsion comprising a blend of EPA/DHA (18 and 21 mg/ml, respectively). Interestingly, this emulsion, already employed in parenteral nutrition, was administered immediately after the onset of MCAO in mice, emphasizing the clinical promise of this therapeutic approach with omega-3 PUFA (Berressem et al., 2016). The results indicated a notable 20% reduction in infarct volume. Specifically, at the striatal level, this treatment led to an 87.5% decrease in glutamate release. Notably, the treatment also resulted in a substantial 49% reduction in the concentration of the pro-inflammatory cytokine interleukin (IL)-6 in the brain. Diets enriched with these omega-3 PUFA have also been reported to be neuroprotective through anti-inflammatory and antioxidant actions after stroke or traumatic brain injury (Wu et al., 2004; Zhang et al., 2010; Pu et al., 2013). The synergistic impact of combined interventions involving both injections and a prophylactically enriched diet has been characterized as notably superior in mitigating tissue atrophy. This approach demonstrates enhanced efficacy in fostering cognitive functions while concurrently preserving the integrity of white matter, which consists of neuronal axons and oligodendrocytes (Jiang et al., 2016; Pu et al., 2016). Oligodendrocyte cells form the myelin sheath around the axons, ensuring better conduction of the nerve signal, and the interaction between oligodendrocytes and axons prevents their degeneration. In the corpus callosum, striatum, and cortex, this treatment promotes oligodendrogenesis (the differentiation and proliferation of oligodendrocytes) and, consequently, decreases demyelination. Treatment also reduces the population of differentiated microglial cells that express a pro-inflammatory phenotype (M1) and promotes their differentiation into the repair phenotype (M2). These results collectively correlate with an attenuation of MCAO-induced sensorimotor deficits and confirm previous data demonstrating oligodendrogenesis in mice treated with a diet enriched in fish oil for three months prior to MCAO (Zhang et al., 2015). These findings collectively highlight the neuroprotective and metabolic benefits of EPA and DHA administration in the context of cerebral ischemia.

EPA-DHA in Clinical Trials.
Following our initial study, which proposed that long-chain omega-3 PUFA, specifically EPA and DHA, may enhance post-stroke recovery by targeting the cerebrovasculature to facilitate vasodilation and improve perfusion (Blondeau et al., 2007), subsequent compelling evidence has affirmed the vasodilatory effects of EPA and DHA, even when integrated into the dietary regimen of patients. A notable study by Kuszewski et al. (2017) showed that the EPA/DHA intake contributes to vasodilation, exerting its effects not only at the systemic level but also within the cerebral vasculature. In particular, the study reported a mean increase of 1.8% in cerebral arterial vasodilation associated with the supplementation of EPA and DHA. This finding highlights the overall potential of omega-3 PUFA to modulate vascular tone and enhance blood flow, particularly within the cerebral circulation.

In addition to PUFA, plants harbor another class of nutraceuticals known as polyphenols, including epigallocatechin 3-gallate (EGCG) and resveratrol found in green tea and grapes, respectively. Distinguished by their structural dissimilarity from PUFAs (Figure 3), polyphenols also possess antioxidant and neuroprotective properties.

**Polyphenols**

*EGCG*

Flavonoids represent a major class of the polyphenol family. Contained in vegetables and fruits, they have multiple biological effects. In plants, they participate in anti-microbial protection and ensure survival in extreme weather conditions. In humans, they have beneficial effects in various pathologies such as cancer, cardiovascular and neurodegenerative diseases (For an overview see: Panche et al., 2016). EGCG is a type of catechin, which is a class of flavonoids, and probably one of the best-known natural flavonoid polyphenols (Figure 3). Highly represented in green tea, EGCG represents 30-40% of the leaves dry weight. The abundant health benefits of this polyphenol have been explored for years. Its antioxidant and anti-inflammatory properties have shown great promise in many areas, including the stroke field.

EGCG is neuroprotective against stroke.
The intravenous administration of EGCG (20 mg/kg) following MCAO induces a noteworthy reduction in infarct volume, concomitant with diminished levels of oxidative stress, expression of caspase-3 and Bax (Hajdukova et al., 2015). The observed protection of EGCG on neuronal damage is postulated to occur through the modulation of apoptosis and the PI3K/Akt signaling pathway. Activation of this pathway leads to upregulation of the anti-apoptotic factor Bcl2 and concurrent inactivation of the pro-apoptotic factor Bax, which is associated with the downregulation of apoptosis-initiating caspase-3 expression (Nan et al., 2018).

EGCG inhibits oxidative stress.
Numerous studies have explored EGCG's ability to contribute to overall redox homeostasis by neutralizing ROS, enhancing cellular antioxidant defenses to counteract oxidative stress, and thereby reducing oxidative damage. Most of the protective effects were attributed to the polyphenolic structure of EGCG, which enables the trapping of free radicals and confers antioxidant properties. To illustrate the robust antioxidant potency of EGCG, in a model of ionizing radiation (radiotherapy) recognized for inducing tissue damage by promoting the generation of free radicals, EGCG demonstrated superior antioxidant efficacy and consequently, cell protection compared to another flavonoid antioxidant, quercetin, as well as vitamin C (Richi et al., 2012). In the stroke context, when administered via arterial pump in the rat MCAO model, EGCG is effective in reducing oxidative stress. This is evident in decreased levels of NO and malondialdehyde (MDA). The notable increase in glutathione peroxidase (GSH-Px) and superoxide dismutase content suggests a potential correlation with inhibiting lipid peroxidation and enhancing antioxidant enzyme activity. This underlines the EGCG interest for promoting antioxidant defenses against stroke-induced oxidative stress (Nan et al., 2018).

EGCG enhances neurogenesis and angiogenesis.
Under physiological conditions, EGCG promotes angiogenesis and the proliferation of neural progenitor cells, thereby improving spatial cognition and learning ability in mice. In stroke, repeated intracerebroventricular EGCG injections every 24 hours for 2 weeks, starting 14 days post-MCAO, enhanced the proliferation of neuronal progenitors within the subventricular region, achieved through the activation of the Akt pathway. These mice exhibit superior cognitive performance in comparison to control mice, indicative of enhanced stroke recovery (Zhang et al., 2017). Beyond its impact on neurons, EGCG also exerts influence on endothelial cells, fostering an elevated rate of endothelial nitric oxide synthase (eNOS) activity. This enzymatic activity, responsible for NO production, induces vasodilation and facilitates reperfusion of brain tissue at the hypoperfused site. Similarly, the injection of 50 mg/kg EGCG in mice post-ischemia stimulates angiogenesis, as shown by an increased number of Ki67/CD31-positive vessels and a higher vascular density in the peri-infarcted area, by upregulating the expression of VEGF and its receptor VEGFR2 (Bai et al., 2017). Despite the evident beneficial properties observed in vivo, it is noteworthy that as of now, no clinical trials have been initiated to investigate the therapeutic potential of EGCG. A principal limit in using EGCG for pharmaceutical purposes lies in its poor bioavailability, notably contributing to therapeutic incompatibilities, particularly in the context of brain targeting. Various strategies aimed at enhancing efficacy were studied, optimizing the administration matrix, utilization of emulsion or nanoparticulate formulations, and the requisite chemical modification of the molecular structure. These approaches are still ongoing to overcome the challenges associated with bioavailability and BBB permeability, thereby advancing the prospects of EGCG as a pharmaceutical agent (for review see: (Furniturewalla and Barve, 2022))

*Resveratrol*
Resveratrol (3,5,4'-trihydroxy-trans-stilbene) is the most studied member of the stilbenoid family and is classified under the polyphenols group (Figure 3). Initially isolated from the roots of white hellebore in 1940, resveratrol is ubiquitously present in over 70 plant species, including but not limited to blueberries, peanuts, coca, as well as in the roots of Japanese and Chinese knotweed plants. Notably, red grapes are a particularly rich source of resveratrol, making it one of the primary bioactive components of red wine, with a content reaching 2 mg/L (Sun et al., 2008; Bertelli and Das, 2009). This polyphenol contributes to the resistance of plants against bacterial and fungal infections. Its biologic and therapeutic implications, especially in cancer and cardiovascular disease, have been the focus of numerous studies (Sadruddin and Arora, 2009) as has its neuroprotective effects ((Richard et al., 2011). In addition, several in vitro, in vivo, and clinical studies revealed its therapeutic potential in the treatment of cerebral ischemia (Berman et al., 2017; Salehi et al., 2018).

Resveratrol has antioxidant properties.
The protective efficacy of resveratrol has been examined across various neuronal conditions associated with insufficient oxygen supply, ranging from hypoxia, characterized by inadequate oxygen reaching cells and causing cellular distress, to ischemia, which entails both oxygen and nutrient deprivation. The antioxidant capability becomes particularly compelling during ischemia/reperfusion scenarios, as the reperfusion phase induces a substantial surge in ROS, triggering oxidative stress, which in turn, precipitates damage to lipids, carbohydrates, proteins, and DNA. Notably, from a nutraceutical standpoint, administering resveratrol via gavage reduces MCAO-induced infarct volumes in mice, even when provided solely during the reperfusion phase (Dong et al., 2008). In rats, intraperitoneal injection of resveratrol reduces infarct volume and decreases oxidative stress as shown by reduced MDA formation and the restoration of superoxide dismutase activity. This neuroprotective effect was further associated with the upregulation of the transcription factor Nrf2 and heme oxygenase-1 (HO-1) (Ren et al., 2011) where Nrf2, upon activation, plays a pivotal role in regulating oxidative pathways by elevating superoxide dismutase and HO-1 levels, leading to a concurrent reduction in ROS content. A similar protective mechanism involving the activation and nuclear translocation of Nrf2 has been observed in rat spinal cord neurons treated with resveratrol during hypoxic injury (Kesherwani et al., 2013). It is noteworthy to emphasize that the in vivo anti-ischemic effects of resveratrol, whether administered via injection or gavage, were achieved through sustained and repeated administration over an extended duration. While the reduction of brain lesions, referred to as neuroprotection, was primarily linked to its inherent antioxidant properties, resveratrol exhibits other protective features that target other detrimental mechanisms induced by ischemia.

Resveratrol reduces glutamatergic excitotoxicity.
Given that glutamate excitotoxicity is a primary contributor to neuronal damage during ischemia, the influence of resveratrol on glutamatergic neurotransmission was examined. In the absence of an ischemic challenge, resveratrol demonstrated a dose-dependent inhibition of glutamate-induced excitation currents in rat hippocampal slices and neurons. This inhibition occurred through interactions with glutamatergic receptors, with a notable emphasis on kainate and NMDA receptors over AMPA receptors (Gao et al., 2006). Growing evidence underscores the pivotal role of astrocytes in neuroprotection, by reducing oxidative stress, secreting neuroprotectants, and modulating synaptic activity. Consequently, astrocytes represent new targets for developing innovative therapeutic strategies for neurodegenerative disorders (Takuma et al., 2004). In astrocytes, resveratrol from 0.1 to 100 μM induced a linear increase in glutamate uptake. However, the efficacy of resveratrol is contingent on its exposure time, and with prolonged exposure, it may exhibit cytotoxic effects (dos Santos et al., 2006). Underscoring potential adverse consequences associated with elevated or prolonged exposures to resveratrol underscores the challenges in promoting the development of clinical trials for this natural product.

Resveratrol shows anti-apoptotic properties.
Rats subjected to intraperitoneal injections of resveratrol (15 or 30 mg/kg) for one week, with an additional dose administered 30 minutes prior to MCAO, exhibit a noteworthy decrease in infarct volume by 28% and 48%, respectively. This neuroprotective effect was linked to a marked reduction in the number of TUNEL-positive cells and cells stained with caspase-3, indicating a reduction in DNA fragmentation (Ren et al., 2011). Moreover, resveratrol enhances the expression of the anti-apoptotic factor Bcl2, while concurrently reducing the expression of the pro-apoptotic factor Bax in the MCAO rats. Consequently, resveratrol may serve as an inhibitor of apoptosis, suggesting its potential to enhance survival and functional outcomes in the post-ischemic neuronal region (Li et al., 2012).

Resveratrol increases angiogenesis.
Exposure of human brain endothelial cells to 5 μM resveratrol for 24 hours promotes their proliferation and their ability to organize into tubes. Indeed, resveratrol activates the PI3K/Akt and ERK pathways, which subsequently activate eNOS. The NO thus produced promotes the expression of VEGF, MMP-2, and MMP-9 (Simao et al., 2012). Taken together, the observations regarding resveratrol indicate its favorable effects on all components of the neurovascular unit, promoting optimal multicellular interactions to maintain neurovascular homeostasis during stroke. Nonetheless, a more detailed examination of how resveratrol influences cell signaling throughout the entire neurovascular unit is imperative.

Resveratrol reduces inflammation and neuroinflammation.
Numerous investigations across diverse peripheral cell types have demonstrated that resveratrol manifests anti-inflammatory effects by inhibiting NF-κB transcriptional activities. Specifically, resveratrol diminishes the expression of the NF-κB subunit RelA/p65 and the translocation of p65 from the cytosol to the nucleus, thereby stabilizing inhibitory I-κB (Manna et al., 2000; Surh et al., 2001). In the context of stroke-induced inflammatory responses in the brain, activated glial cells, particularly microglia and astrocytes, play a pivotal role. In murine microglia and astrocytes, the in vitro application of resveratrol produces a reduction in LPS-induced expression of pro-inflammatory cytokines, including TNF-α, IL-6, MCP-1, and iNOS. This reduction was achieved by inhibiting NF-κB activation in both cell types, suggesting a marked impact of resveratrol on neuroinflammation. Furthermore, these anti-inflammatory effects appear to be associated with the SIRT1/SOCS1 pathway (Lu et al., 2010). Similarly, intraperitoneal injection of resveratrol before MCAO in rats decreases TNFα and IL-1β levels and downregulates the TLR4 signaling pathway, thereby attenuating inflammation and reducing the infarct size (Lei et al., 2019). Taken together, the preclinical studies suggest that resveratrol may help control stroke-induced inflammation, BBB disruption, and brain damage. Recently, evidence has emerged indicating that the impact of resveratrol on neuroinflammation, achieved by reducing microglial activation, may be mediated through the enhancement of intestinal microbiota (He et al., 2022).

Resveratrol modulates the bowel-brain axis.
While the limited permeability across the BBB of resveratrol may limit the action of its systemic administration on local brain processes, resveratrol's benefits for the brain may emerge from various physiological effects exerted throughout the body in the context of stroke. One emerging area of interest in understanding these systemic effects is the role of the gut-brain axis in stroke pathology (Benakis and Liesz, 2022). The gut-brain axis refers to the bidirectional communication between the gastrointestinal tract and the central nervous system. Emerging evidence suggests that the gut microbiota, the diverse community of microorganisms residing in the gastrointestinal tract, can influence neurological functions and potentially play a role in the outcomes of neurological disorders, including stroke (Benakis and Liesz, 2022). Resveratrol intake may interact with the gut microbiota and modulate its composition. This modulation can, in turn, influence the production of metabolites and signaling molecules that affect the central nervous system (Ochoa-Reparaz et al., 2011; Chen et al., 2016b; Bird et al., 2017; Kim, 2024). Furthermore, the intraperitoneal administration of resveratrol to mice during the three days following MCAO results in an elevation of Th2 and T-regulatory lymphocyte levels in the gut. These lymphocytes actively secrete anti-inflammatory cytokines, such as IL-4 and IL-10, leading to a consequential decrease in the secretion of pro-inflammatory cytokines, TNFα and IFNγ, by other lymphocytes. This comprehensive modulation contributes to the reduction of both intestinal vascular and cerebral permeability, ultimately resulting in decreased infarct volume. Considering the pivotal role of the intestinal microbiota in lymphocytic differentiation and maturation, the neuroprotective effects of resveratrol may be attributed to its potential to promote the proliferation of beneficial bacteria and/or reduce the abundance of pathogenic bacteria within the intestine, as proposed by Dou et al. (2019).

Clinical use of resveratrol.
Given the widespread over-the-counter availability of resveratrol as a dietary supplement and the limited clinical data, the medicinal and clinical utilization of resveratrol remains a dynamically evolving area of research. Predominantly, clinical trials investigating resveratrol have concentrated on cancer, neurological disorders, cardiovascular diseases, diabetes, non-alcoholic fatty liver disease, and obesity (Berman et al., 2017). Nevertheless, within the few clinical trials assessing the neuroprotective effects of resveratrol, particularly in conditions such as Alzheimer's disease, some studies have suggested potential benefits associated with resveratrol supplementation (Turner et al., 2015; Zhu et al., 2018). In the context of stroke, a limited number of trials have sought to elucidate the safety and efficacy of resveratrol in patients with acute ischemic stroke, yielding varied outcomes. Despite oral administration of resveratrol at a daily dose of 500 ± 10 mg, distributed in three 170 mg divided doses over 30 consecutive days post-stroke, does not improve the functional recovery measured by the NIHSS, Modified Rankin Score, and Barthel index, resveratrol may hold significant interest as an adjuvant with r-tPA treatment, potentially extending its therapeutic window. Notably, in patients receiving delayed r-tPA treatment, co-administration of resveratrol significantly enhances treatment outcomes, as evidenced by improved NIHSS scores. This observed improvement is attributed to resveratrol-induced reductions in plasma levels of both MMP-2 and MMP-9 (Chen et al., 2016a). However, the clinical application of resveratrol as a therapeutic agent faces several constraints, notably its principal limitation: low bioavailability and poor solubility, tempering enthusiasm for its widespread clinical use (Walle, 2011; Salehi et al., 2018). The current state underscores the imperative for further rigorous clinical investigations to delineate its efficacy, optimal dosage, and overall effectiveness before definitive conclusions can be drawn regarding its broad clinical applicability (Chachay et al., 2011).

**Conclusion**

In summary, this review underscores the significance of nutraceuticals, such as ALA, EPA, DHA, EGCG, and resveratrol, in mitigating the detrimental effects of cerebral ischemia. These natural compounds are increasingly recognized for their potential in both preventing and treating stroke. Broadly, they exhibit antioxidant, anti-inflammatory, and anti-apoptotic properties, acting comprehensively on the neurovascular unit. At the neuronal level, these compounds modulate excitotoxicity, facilitate neurogenesis and synaptogenesis, thereby preserving neuronal functionality and the integrity of brain tissue. On the vascular front, they stimulate angiogenesis and vasodilation, promoting brain reperfusion and contributing to restored function post-stroke.

Importantly, ALA, EPA, DHA, and resveratrol have demonstrated some efficacy in human studies, suggesting their potential as key elements in novel preventive and therapeutic strategies against stroke. Given their presence in commonly consumed foods, incorporating these nutraceuticals into daily dietary habits, especially for predisposed individuals, could serve as a robust foundation for reducing the risk of stroke and other serious health issues. This dietary adaptation holds promise for enhancing overall health and well-being.

**Acknowledgments**

This work was supported by the CNRS. The authors thank all past and present team members, with special appreciation to our former student, Pavel Kotchetkov, for inspiring this review and providing skillful technical assistance, as well as to our collaborators who contributed to the data discussed herein.

**Conflict of Interest**

The authors declare no competing financial interests. NB , who serves on the Editorial Board of Conditioning Medicine did not participate in the editorial review of this manuscript at any level.